\documentclass[fleqn,twoside,twocolumn,nofootinbib,showkeys]{revtex4} % Specifies the document class %,unsortedaddress
\usepackage[nocpr,nopacs]{ujp} % \usepackage[cyr]{ujp} for cyrillic, \usepackage[web]{ujp} for web
\begin{document}
\title[About Conditions of Spatial Collapse ]%колонтитул
{ABOUT CONDITIONS OF SPATIAL COLLAPSE \\ IN AN INFINITE SYSTEM OF BOSE PARTICLES}%

\author{B.E.~Grinyuk}%1 автор
\affiliation{Bogolyubov Institute for Theoretical Physics, Nat.  Acad. of Sci. of Ukraine}%
\address{14-B, Metrolohychna Str., Kyiv 03143, Ukraine}%
\email{bgrinyuk@bitp.kiev.ua}%e-mail 1

\author{\fbox{K.A.~Bugaev\!}}
\affiliation{Bogolyubov Institute for Theoretical Physics, Nat.  Acad. of Sci. of Ukraine}%
\address{14-B, Metrolohychna Str., Kyiv 03680, Ukraine}%
\email{bgrinyuk@bitp.kiev.ua}%e-mail 1

\udk{539}  \razd{\seci}

\autorcol{B.E.\hspace*{0.7mm}Grinyuk, K.A.\hspace*{0.7mm}Bugaev}

\setcounter{page}{1024}%

\begin{abstract}
Using the variational principle, we show that the condition of
spatial collapse in a Bose gas is not determined by the value of the
scattering length of the interaction potential between particles
contrary to the result following from the Gross--Pitaevskii
equation, where the collapse should take place at a negative
scattering length.
\end{abstract}

 \maketitle

\section{Introduction}

The spatial collapse of an infinite system of interacting Bose
particles was commonly analyzed  within the approach based on the
Gross--Pitaevskii \cite{R1,R2,R3,R4} equation  (see also the latest
studies \cite{R5}).\,\,An obvious result of this analysis lies in
the fact that, under the condition $a<0$ \cite{R6} (where $a$ is the
two-particle scattering length determining the sign of non-linear
term of the equation), one has a collapse in the system.

In the present paper, starting from the initial Hamiltonian of a
system of $N$ Bose particles and using the variational principle,
we show that the spatial collapse of the Bose system at
$N\rightarrow\infty$, generally speaking, is not determined by the
two-particle scattering length value.

\section{Statement of the Problem}

We consider a system of $N$ identical interacting Bose particles
of mass $m$ with the Hamiltonian
%1
\begin{equation}\label{E1}
\hat{H}=\sum_{k=1}^{N}\frac{\hat{\mathbf{p}}_{k}^{2}}{2m}+\sum_{n>k=1}^{N}V\left(\left|\bf{r}_{n}-\bf{r}_{k}\right|\right)\!,
\end{equation}
where the pairwise potential depends on the distance between
particles. In the present paper, we consider a rather wide class
of potentials obeying the inequality
%2
\[
V\!\left(r\right)<V_{0}\left(r\right)\equiv
\]\vspace*{-7mm}
\begin{equation}\label{E2}
\equiv V_{01}\,\exp\left(\!-\left(r/r_{01}\right)^{2}\!\right)-
V_{02}\,\exp\left(\!-\left(r/r_{02}\right)^{2}\!\right)
\end{equation}
with $r_{01}<r_{02}$ and $V_{01}\geq 0$, $V_{02}\geq 0$.\,\,If, for
the ``reference'' potential $V_{0}\left(r\right)$ with a definite
set of parameters, one obtains the collapse in the Bose system
(\ref{E1}), then, for any potential
$V\!\left(r\right)\!<\!V_{0}\left(r\right),$ the collapse will
obviously take place even more.\,\,The ``reference'' potential is
chosen in the form (\ref{E2}) for the convenience in order to
calculate some integrals in an explicit form (see below).

Now, we are going to carry on a rather simple variational estimation
of the ground state energy of the system of Bose particles with the
Hamiltonian (\ref{E1}) and to find a sufficient condition for the
spatial collapse of the system to exist.\,\,It should be stressed
that the condition to be obtained will be not a necessary criterion,
but only a sufficient one.\,\,In principle, one can carry out a more
accurate variational estimation and obtain a more refined sufficient
condition.

\section{Variational Estimation\\ with One-Particle Trial Functions}

As is well-known (see, e.g., \cite{R7}), the ground-state energy
$E_{0}$ of any quantum system does not exceed the average
%3
\begin{equation}\label{E3}
E_{0}\leq\frac{\left\langle \Psi\right|\hat{H}\left|
\Psi\right\rangle}{\left\langle \Psi | \Psi\right\rangle},
\end{equation}
where $\Psi$ is an arbitrary (``trial'') function.\,\,The sign of
equality is valid, if $\Psi$ is the exact solution of the
Schr\"{o}dinger equation for the ground state of the
system.\,\,Fur\-ther, we will consider the Hamiltonian $\hat{H}_{0}$
with a ``reference'' potential $V_{0}$ of the special
form~(\ref{E2}).

Consider the trial function $\Psi$ in the simplest form of a product
of one-particle functions:\vspace*{-1mm}
%4
\[
\Psi\left(\mathbf{r}_{1},\mathbf{r}_{2},...,
\mathbf{r}_{N}\right)=\prod_{k=1}^{N}\exp\left(\!-\left(\mathbf{r}_{k}/R\right)^{2}\!\right)\equiv
\]\vspace*{-6mm}
\begin{equation}\label{E4}
\equiv\exp\left(\!-\frac{1}{R^{2}}\sum_{k=1}^{N}\mathbf{r}_{k}^{2}\!\right)\!\!,
\end{equation}\vspace*{-3mm}

\noindent where $R$ characterizes the size of the system.\,\,The
choice of a trial function in the form (\ref{E4}), as well as the
choice of a ``reference'' potential, enables one to calculate
integrals explicitly.\,\,In particular, the normalization integral
results in\vspace*{-1mm}
%5
\begin{equation}\label{E5}
\left\langle \Psi |
\Psi\right\rangle=\left(\!\frac{\pi}{2}\!\right)^{\!\!\frac{3N}{2}}R^{3N}\!\!.
\end{equation}\vspace*{-4mm}

\noindent The average of the Hamiltonian consists of two
terms.\,\,The first one is the matrix element of the kinetic
\mbox{energy}\vspace*{-1mm}
%6
\[
\left\langle \Psi
\right|\sum_{k=1}^{N}\frac{\hat{\mathbf{p}}_{k}^{2}}{2m}
\left|\Psi\right\rangle\equiv-\frac{\hbar^{2}}{2m}\sum_{k=1}^{N}\left\langle
\Psi \right|\triangle_{k}\left|\Psi\right\rangle=
\]\vspace*{-6mm}
\begin{equation}\label{E6}
=N\frac{3\hbar^{2}}{2mR^{2}}\left(\!\frac{\pi}{2}\!\right)^{\!\!\frac{3N}{2}}\!\!R^{3N}\!\!,
\end{equation}\vspace*{-4mm}

\noindent and the second one is the potential energy matrix
element:\vspace*{-1mm}
%7
\[
\left\langle \Psi
\right|\sum_{n>k=1}^{N}V_{0}\left(\left|\mathbf{r}_{n}-\mathbf{r}_{k}\right|\right)
\left|\Psi\right\rangle=
\]\vspace*{-6mm}
\begin{equation}\label{E7}
=\frac{N\!\left(N\!-\!1\right)}{2}\,\frac{1}{R^{3}}\left(V_{01}\rho_{01}^{3}-
V_{02}\rho_{02}^{3}\right)\left(\!\frac{\pi}{2}\!\right)^{\!\!\frac{3N}{2}}\!\!R^{3N}\!\!,
\end{equation}\vspace*{-4mm}

\noindent where $\rho_{0j}\equiv
r_{0j}\left(\!1+r_{0j}^{2}/R^{2}\!\right)^{\!\!-1/2}$,
$j=1,2$.\,\,If parameter $R$ is essentially greater than the radius
of forces, $R\gg r_{0j}$, one has $\rho_{0j}\rightarrow r_{0j}$.

Using integrals (\ref{E5})--(\ref{E7}) in the explicit form, we obtain
the following estimation for the ground-state energy of the
Hamiltonian $\hat{H}_{0}$:\vspace*{-1mm}
%8
\begin{equation}\label{E8}
E_{0}\leq
N\frac{3\hbar^{2}}{2mR^{2}}+\frac{N\!(N-1)}{2}\,\frac{1}{R^{3}}(V_{01}\rho_{01}^{3}-
V_{02}\rho_{02}^{3}),
\end{equation}\vspace*{-4mm}

\noindent or, for a finite $R$, but $R\gg r_{0j}$, and divided by
the number of particles,\vspace*{-1mm}
%9
\begin{equation}\label{E9}
\frac{E_{0}}{N}\leq
\frac{3\hbar^{2}}{2mR^{2}}+\frac{N\!-\!1}{2}\,\frac{1}{R^{3}}(V_{01}r_{01}^{3}-
V_{02}r_{02}^{3}).
\end{equation}
It obviously follows from estimation (\ref{E9}) that, under the
condition\vspace*{-1mm}
%10
\begin{equation}\label{E10}
V_{01}r_{01}^{3}- V_{02}r_{02}^{3}<0,
\end{equation}
the energy per one particle (\ref{E9}) tends to $-\infty$, as
$N\rightarrow \infty$.\,\,In addition, the less the parameter $R$,
the lower is the energy due to the $\sim$$ R^{-3}$ dependence of the
potential energy as compared to the $\sim$$ R^{-2}$ dependence of
the kinetic energy.\,\,Thus, one has a spatial collapse of the
system of Bose particles to a small area of the order of the radius
of forces (or a little bit greater) with an infinite density and an
infinite negative energy even calculated per one particle.

\section{Discussion}
Since the integral over the ``reference'' potential
$V_{0}\left(r\right)$ equals\vspace*{-1mm}
%11
\begin{equation}\label{E11}
\int V_{0}\left(r\right)d\mathbf{r}=\pi^{3/2}(V_{01}r_{01}^{3}-
V_{02}r_{02}^{3}),
\end{equation}
one can use the condition\vspace*{-1mm}
%12
\begin{equation}\label{E12}
\int V_{0}\left(r\right)d\mathbf{r}<0
\end{equation}
instead of (\ref{E10}).\,\,It is well known that the two-particle
scattering length $a$ is expressed through the scattering amplitude
at the zero transferred momentum as $a=-\left(k
\cot\delta\left(k\right)\right)^{-1}|_{k\rightarrow 0}$, and, in the
Born approximation, $a_{\rm Born}$ is proportional to the integral
over the potential \cite{R7}\vspace*{-1mm}
%13
\begin{equation}\label{E13}
a_{\rm Born}=\frac{\mu}{2\pi \hbar^{2}}\int
%V_{0}\left(r\right)d\mathbf{r}=\frac{4\pi^{2}\mu}{\hbar^{2}}\tilde{V}\left(0\right),
V_{0}\left(r\right)d\mathbf{r},
\end{equation}
where $\mu$ is the reduced mass of two particles ($\mu=m/2$ for
identical particles).\,\,Thus, condition (\ref{E12}) can be simply
rewritten as $a_{\rm Born}<0$, where the scattering length in the
Born approximation is calculated for the ``reference'' potential
$V_{0}\left(r\right)$.\,\,Mo\-reover, it can be shown for any
short-range potential $V\left(r\right)$ that the condition like
(\ref{E12}) or $a_{\rm Born}<0$ is valid, if the parameter $R$ in
function (\ref{E4}) is essentially greater than the radius of forces
(but remains to be finite in order to we can speak about spatial
collapse).

It is important to emphasize, that more accurate variational
estimations will never lead us to the condition of collapse like
$a<0$, which follows from the Gross--Pitaevskii equation (where $a$
is the exact scattering length).\,\,Really, let us start with some
``reference'' potential $V_{0}\left(r\right)$ giving an
infinitesimal, but already negative integral (\ref{E12}).\,\,As a
result, we already have the collapse of the system.\,\,Let us add an
attraction term  of the intensity $-g$ $(g>0)$, and with the radius
$r_{02}$, to this potential.\,\,Thus, we obtain a potential
$V\left(r\right)$ in the form:\vspace*{-1mm}
%14
\begin{equation}\label{E14}
V(r)\equiv V_{0}(r)-g\exp\,(-(r/r_{02})^{2}).
\end{equation}

%Figure
\begin{figure}
\vskip1mm
\includegraphics [width=\column] {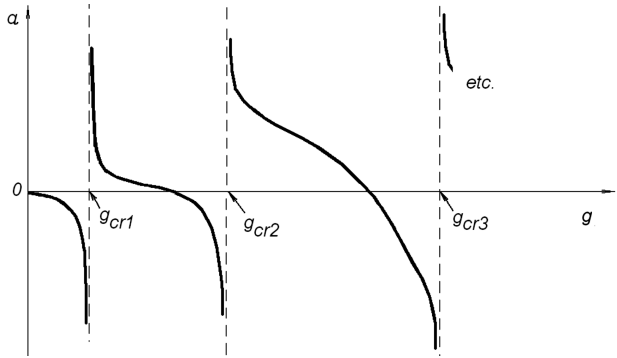}
\vskip-2mm\caption{A schematic typical dependence of the
scattering length $a$ on the intensity of attraction $g$}
\end{figure}

\noindent Since we have $V(r)<V_{0}(r)$, the collapse in the system
of Bose particles, all the more, will take place at any
$g>0$.\,\,But a dependence of the scattering length $a$ on $g$ is
known to be like that shown in Figure.\linebreak This dependence can
be verified using the Lippmann--Schwinger integral equation
\cite{R8} for the $t$-matrix.\,\,To find the scattering length, the
variable phase approach to the potential scattering can also be used
\cite{R9,R10}, where a first-order differential equation is to be
solved.\,\,In Figure, we depict a typical dependence of the
scattering length on the intensity $g$ for the potential
(\ref{E14}).\,\,At $g=0$, one has the almost zero, but negative
scattering length for this potential.\,\,Near the critical values of
the attraction intensity $g$ (where new two-particle bound states
appear), the scattering length is known to have non-removable
discontinuity, and it changes its value from minus to plus
infinity.\,\,One can easily see that the scattering length may have
any value and sign, but the collapse of the system is still present
and even becomes stronger and stronger with an increase of
$g$.\,\,Note that, at the same time, the integral over the potential
and, thus, the scattering length in the Born approximation $a_{\rm
Born}$ remains
 to be negative, as the constant $g$ grows.\,\,Thus,
the condition $a_{\rm Born}<0$ may serve to be a sufficient
condition for the collapse to take place.\,\,But we stress that it
is not the necessary condition: more accurate variational
estimations may show the existence of the Bose system collapse even
in cases where $a_{\rm Born}>0$.\looseness=1

\section{Conclusions}

To summarize, we note that the scattering length $a$ of the
two-particle interacting potential cannot serve to be the value
determining the presence or absence of a collapse in an infinite
system of interacting Bose particles, in the contrary to the
conclusions following from the Gross--Pitaevskii equation.\,\,This
statement is substantiated with the use of the variational principle
with the one-particle approximation for trial functions.\,\,The
sufficient condition obtained in the present paper for the collapse
in the Bose system of particles $a_{\rm Born}<0$ can be improved,
and, thus, it is not necessary condition.

%\section{Acknowledgements}

%This work is partially supported by the National Academy of
%Sciences of Ukraine, Project No.\,\,0117U000237, and by the Project
%No.\,\,CC-10-2021.
\vskip3mm \textit{This work was supported by the National Academy of
Sciences of Ukraine, Contract No.\,\,CC-10-2021.}

\vskip-5mm \rezume {Б.Є.\,Гринюк, К.О.\,Бугаєв}{ЩОДО УМОВ
ПРОСТОРОВОГО КОЛАПСУ \\ В НЕСКІНЧЕННІЙ СИСТЕМІ БОЗЕ-ЧАСТИНОК}{На
основі варіаційного принципу показано, що умова просторового колапсу
в бозе-газі не визначається величиною довжини розсіяння для
потенціалу взаємодії між частинками, на відміну від результату, який
випливає з рівняння Ґроса--Пітаєвського і означає наявність колапсу
у випадку від'ємної довжини
розсіяння.}{\textit{К\,л\,ю\,ч\,о\,в\,і\, с\,л\,о\,в\,а:}
бозе-система, просторовий колапс. }

\end{document}